# Nulling at short wavelengths: theoretical performance constraints and a demonstration of faint companion detection inside the diffraction limit with a rotating-baseline interferometer

E. Serabyn,[1] B. Mennesson,[1] S. Martin,[1] K. Liewer,[1] and J. Kühn[2]

[1]Jet Propulsion Laboratory, California Institute of Technology, Pasadena, CA, 91109, USA

[2]University of Bern, Center for Space and Habitability, Gesellschaftsstrasse 6, 3012, Bern, Switzerland

## Abstract

The Palomar Fiber Nuller (PFN) is a rotating-baseline nulling interferometer that enables high-accuracy near-infrared (NIR) nulling observations with full azimuth coverage. To achieve NIR null-depth accuracies of several x $10^{-4}$, the PFN uses a common-mode optical system to provide a high degree of symmetry, single-mode-fiber beam combination to reduce sensitivity to pointing and wavefront errors, extreme adaptive optics to stabilize the fiber coupling and the cross-aperture fringe phase, rapid signal calibration and camera readout to minimize temporal effects, and a statistical null-depth fluctuation analysis to relax the phase stabilization requirement. Here we describe the PFN's final design and performance, and provide a demonstration of faint-companion detection by means of nulling-baseline rotation, as originally envisioned for space-based nulling interferometry. Specifically, the $K_s$-band null-depth rotation curve measured on the spectroscopic binary η Peg reflects both a secondary star 1.08 ± 0.06 x $10^{-2}$ as bright as the primary, and a null-depth contribution of 4.8 ± 1.6 x $10^{-4}$ due to the size of the primary star. With a 30 mas separation at the time, η Peg B was well inside both the telescope's diffraction-limited beam diameter (88 mas) and typical coronagraphic inner working angles. Finally, we discuss potential improvements that can enable a number of small-angle nulling observations on larger telescopes.

## 1. Introduction

Faint emission very near bright stars can be made easier to detect by using nulling interferometry to suppress the starlight arriving at multiple collecting apertures. Indeed, stellar nulling on a rotating space-based interferometer was originally proposed as a means of detecting the thermal emission from Earth-like exoplanets (Bracewell 1978; Angel 1990; Leger et al. 1996; Mennesson & Mariotti 1997). However, detecting astronomical companions using baseline rotation, as opposed to Earth rotation, has yet to be demonstrated, even from the ground, as baseline rotation is generally not feasible with separate ground-based telescopes. Indeed, even the co-mounted LBTI apertures (Hinz et al. 2016) rely on Earth rotation to provide relative baseline-source rotation. However, as demonstrated by the Palomar Fiber Nuller (PFN), interferometric baselines implemented across the pupil of a large single-aperture telescope can be rotated simply by rotating an image of the telescope pupil (Mennesson et al. 2006; Serabyn & Mennesson 2006; Martin et al. 2008; Mennesson et al. 2010; Serabyn et al. 2010). This enables active baseline rotation to be made use of in ground-based nulling demonstrations and observations.

One advantage of nulling interferometry across a large telescope pupil (i.e., "cross-aperture nulling") is that it can provide high-contrast observations at angles significantly smaller than those attainable with either classical full-aperture coronagraphs or dual-star interferometry, both of which typically operate at angles beyond $\lambda/D$, where $\lambda$ is the observation wavelength and D the aperture diameter (Guyon et al. 2006; Shao and Colavita 1996; Lacour et al. 2019). On the other hand, interferometry can generally reach angles $< \lambda/2D$, i.e., well within the diffraction core of the stellar point spread function (PSF). Of course, interferometers that operate within the PSF core necessarily have more modest contrast capabilities than coronagraphs and interferometers that observe further from the primary star, even with identical raw stellar suppressions. As a result, prior to the advent of the Palomar Fiber Nuller (PFN), ground-based astronomical interferometry had been limited at best to visibility (or null depth) accuracies of about $10^{-2}$ to $10^{-3}$ (Absil et al. 2011), far from the contrast levels needed for exoplanet detection, $10^{-6}$ - $10^{-7}$ for terrestrial exoplanets in the mid-infrared (MIR), and $\sim 10^{-4}$ for the innermost known Hot Jupiters in the near-infrared (NIR). (The new GRAVITY dual-star interferometer [Lacour et al. 2019] shows similar intrinsic fringe visibilities, but reaches better contrasts by observing at several $\lambda/D$ from the star.) A more favorable case within reach of small-angle interferometric observations, i.e., within the $\lambda/D$ diffraction limit, is that of very young, still self-luminous or accreting exoplanets, with potential contrasts $\sim 10^{-2} – 10^{-3}$ (Huelamo et al. 2011, Kraus & Ireland 2012, Haffert et al. 2019). Brown dwarf companions can also show contrasts up to $\sim 10^{-3}$, implying that high-contrast nulling can potentially be used to search for brown dwarf companions to nearby stars at very small separations, i.e., $< \lambda/D$, including, for example, more-massive long-term-trend radial-velocity candidates such as those recently imaged beyond a few $\lambda/D$ (Crepp et al. 2016; Ryu et al. 2016). To date nulling has mostly been employed to observe MIR dust emission (Hinz et al. 1998; Milan-Gabet et al. 2011; Mennesson et al. 2014; Defrere et al. 2015, 2016; Ertel et al. 2018), but NIR nulling interferometry has the potential to provide better contrast than visibility measurements, and so could allow exploration of, e.g., the small NIR "visibility deficits" seen around several nearby main-sequence stars (Absil et al. 2013, Ertel et al. 2014, Nunez et al. 2017), which may be due to hot dust reservoirs at small angles. High contrast NIR nulling observations at angles $< \lambda/D$ could therefore potentially enable high-contrast observations of a number of interesting faint target categories.

The PFN was built with several goals in mind. Our primary goal was to demonstrate companion detection by means of baseline rotation, as originally envisioned for proposed space-based interferometers such as the Darwin interferometer (Leger et al. 1996) and the Terrestrial Planet Finder Interferometer (TPF-I; Beichman et al 1999). The second was to demonstrate companion detection at angles much smaller than full-aperture coronagraphy can reach, i.e., within the stellar PSF's diffraction core. The third was to reach deeper contrasts ($\sim 10^{-3}$ to $10^{-4}$) than those typically provided by interferometric visibility measurements. Fourth, we aimed to extend the use of nulling techniques from the MIR into the shorter-wavelength NIR region, where the dominant noise source is phase instability instead of thermal background fluctuations. The PFN had earlier demonstrated the latter three goals (Hanot et al. 2010, 2011; Mennesson et al 2011a, 2011b), and here we present a clear demonstration of faint companion detection within the PSF core using baseline rotation. At the same time, even with its modest 3.4 m baseline, the PFN was able to begin acquiring interesting high contrast observations, including stellar diameter measurements (Hanot et al. 2011, Mennesson et al. 2011a), deep limits to hot exozodiacal dust around Vega (one of the visibility-deficit/hot-dust stars; Mennesson et al. 2011b), and the detection of hot dust within 4 – 10 AU of the

Herbig star AB Aur (Kühn et al. 2015). PFN observations of the brightest visibility-deficit/hot-dust stars will be addressed separately.

This paper begins with a discussion of the essential requirements for high contrast nulling in the NIR, and describes how these considerations led to the PFN's implementation and data reduction approaches. It next briefly describes some of the observations which an NIR nuller can enable. We then present nulling observations of the spectroscopic binary η Peg which provide a clear demonstration of the use of nulling baseline rotation for both companion detection and stellar diameter measurement. We end with a discussion of the PFN's current limitations and the potential for improvement, and of the advances that nullers on larger telescopes might be able to achieve. Indeed, the potential of nulling observations on larger telescopes is significant, as the lengths of cross-aperture baselines on 30-40 m telescopes would be in a regime previously reachable only by means of separated-aperture long-baseline interferometry (LBI), thereby bringing the innermost Hot Jupiters within reach. Finally, a pair of appendices provide further detail on two topics: the main reason for carrying out nulling instead of visibility measurements (lower noise at the fringe minimum), and the PFN's final instrumental configuration.

## 2. High-accuracy cross-aperture NIR nulling

**2.1. Inner working angle (IWA).** Interferometric stellar nulling between a pair of sub-apertures located within a large telescope pupil is based on centering an achromatic destructive interference fringe across the stellar disk, which suppresses the starlight relative to off-axis emission. Defining, as usual, the interferometric IWA as the half-power point of the central fringe, the IWA for a sub-aperture nuller of baseline b is then λ/4b, with of course b < D, where D is the telescope pupil diameter. The term "inner working angle" should however not be taken too literally, because observations can in fact be undertaken at angles smaller than the fringe half power point, but with lower transmission. With a symmetrically-located pair of sub-apertures of diameter d within the telescope pupil, the maximum possible baseline between their centers, $b_{max}$, is $b_{max}$ = D-d, and the smallest possible IWA is then

$$\text{IWA}_{min} = \frac{\lambda}{4(D-d)}. \tag{1a}$$

There are two opposing drivers on the sub-aperture diameter: a larger d brings a higher signal-to-noise ratio (SNR), enabling interferometry on fainter stars, while a smaller d enables observations both closer to the star (Eqn. 1a) and out to a larger "outer working angle" (= λ/2d) due to the larger single-mode sub-aperture beam width (λ/d). A useful SNR for NIR interferometry requires subaperture diameters of at least 1 m or so, so for the PFN we opted for the largest subapertures that could fit between the outer edges of the primary mirror and the secondary obscuration, i.e., d = 1.5 m. To maximize the SNR, the pupil subapertures were also typically elliptically elongated in the direction perpendicular to the baseline. For the PFN case (D = 5 m and d = 1.5 m), Eqn. 1a gives IWA = 0.36 λ/D, i.e., 32 mas in the $K_s$ band. On the other hand, for a 30 m telescope, similarly-sized subapertures imply d << D, for which Eq. 1a simplifies to

$$\text{IWA}_{min} \approx \frac{\lambda}{4D}. \tag{1b}$$

Because D - d ranges only from 0.7 D to D between these two cases, the resultant IWAs are actually similar in λ/D units in both cases, ~ 0.3 λ/D. Such IWAs are significantly smaller than the full width at half maximum (FWHM) of the full telescope's diffraction-limited PSF (≈ λ/D), and a factor of ~ 3 - 10 smaller than IWAs of modern well-corrected coronagraphs (Guyon et al. 2006). Of course, this comes with a throughput penalty, as cross-aperture nulling may not make use of the full telescope aperture. On the other hand, coronagraphs that do use the full telescope aperture typically cannot reach inside λ/D, one potential exception being the recently proposed vortex fiber nuller (Ruane et al. 2018).

However, while sub-aperture nullers can probe closer to stars than coronagraphs can, so can other types of cross-aperture interferometry, such as aperture masking interferometry (Tuthill et al. 2000). So, why null? As Appendix 1 shows, the main reason is that operating at null minimizes noise. In addition, visibility measurements, which can provide closure phases for three or more baselines, provide access to asymmetric source structures (whether companions or dust clumps), while the visibility amplitudes, which provide access to centro-symmetric dust structures as well, tend to have limited accuracies (~ $10^{-2}$ - $10^{-3}$). Being intensity-based, nulling can detect both symmetric and asymmetric source structures, making nulling and aperture masking complementary in that regard.

**2.2. Null depth occurrence rates.** To exploit the small-angle regime effectively, a nuller must minimize stellar leakage error terms to provide deep and stable rejection of starlight. This requires high degrees of instrumental symmetry and stability (Serabyn 2000; Martin et al. 2003; Serabyn et al. 2012; Serabyn 2020). Cross-aperture nulling within a single telescope pupil has the advantage that a largely common-mode optical system can be employed, which allows for a relatively symmetric optical beam train. In the NIR, due to the reduced thermal background compared to the ground-based MIR regime where the Keck and LBTI nullers operate (Colavita et al. 2009; Hinz et al. 2016), phase fluctuations are expected to dominate the noise, implying that maintaining the null phase is the highest priority. Cross-aperture phase stability sufficient for useful nulling at wavelengths as short as the NIR is now possible due to three factors: 1) ExAO systems on large telescopes that can stabilize the phase across a telescope aperture to ~ 100 nm rms can essentially act as a cross-aperture interferometric fringe tracker, 2) high-quality NIR single-mode (SM) fibers can ease the pointing and wave-front error requirements, as has already been shown for the case of NIR visibility measurements (Coude du Foresto et al. 1998; Mennesson et al. 2002), and 3) the introduction of the nulling self-calibration (NSC) algorithm for null depth determination in the presence of residual fluctuations can greatly relax the phase stability requirements (Hanot et al. 2010, 2011; Mennesson et al. 2011a, 2011b, 2013, 2016; Serabyn 2020). We now briefly discuss these aspects.

The primary goal of any nulling interferometer is to maximize the amount of time spent at or near the null fringe minimum. In the simplest case of small phase deviations $\phi$ from the null phase of $\pi$ radians, what is measured is a time series of null depth values, $N_m$, given by

$$N_m = N_a + N_i = N_a + \frac{\phi^2}{4}, \qquad (2)$$

where $N_a$ is the true astrophysical null depth, and $N_i$ is the instrumental null-depth contribution, given by

$$N_i = \frac{\phi^2}{4}. \qquad (3)$$

For the simple case of Gaussian phase fluctuations with a root-mean-square (rms) deviation of $\sigma_\phi$ about the bottom of the null fringe, the probability, $p(N_i < N_o)$, that $N_i$ is less than some specified null depth, $N_o$, is given by the error function (Bevington 1969):

$$p(N_i < N_o) = p(|\phi_i| < |\phi_o|) = \mathrm{erf}\left(\frac{\phi_o}{\sqrt{2}\sigma_\phi}\right) \tag{4a}$$

where the error function integrates the Gaussian errors between the opposite phase errors, $\pm\phi_o$, that yield the same instrumental null depth, $N_o = \phi_o^2/4$. The rms phase error can be converted to the rms pathlength error, $\sigma_x$, using $\phi = 2\pi x/\lambda$, with x the pathlength error, giving

$$p(N_i < N_o) = p(|x| < |x_o|) = \mathrm{erf}\left(\frac{\lambda}{\pi\sigma_x}\sqrt{\frac{N_o}{2}}\right). \tag{4b}$$

Fig. 1 plots the time fraction expected to be spent below a given null level for representative AO and ExAO cases, i.e., for Strehl ratios of 0.6 and 0.95, respectively. From this Figure, it is evident that the improvement in the typical null depths provided by an upgrade from a first to a second generation AO system is roughly an order of magnitude. Specifically, in the $K_s$ band at ~ 2.15 µm, for a typical ExAO pathlength correction level of $\sigma_x$ ~ 100 nm, the time fraction spent below $N_o = 10^{-1}$, $10^{-2}$, $10^{-3}$, and $10^{-4}$, is 97%, 51%, 17% and 5.5%, respectively. Even with ExAO wavefront correction, only a small time fraction is thus actually spent at very deep null levels. Although some finer stabilization of the piston phase difference between subapertures is possible on bright stars because the nulling apertures are much larger than typical wavefront-sensor subapertures, we instead turned to a statistical analysis of the null-depth fluctuations to obviate the need for further stabilization, as is elaborated in the next section.

**2.3. Null depth statistics**. In the presence of phase fluctuations, the measured null depth will also be a fluctuating quantity (Eqn. 2). We therefore next consider its time average, which from Eqn. 2 is given by

$$\overline{N_m} = N_a + \frac{\overline{\phi^2}}{4}. \tag{5}$$

The quadratic dependence of null depth fluctuations on phase errors, which implies positive-definite null depth fluctuations, thus leads to a positive bias in the measured null depth. If the mean phase is zero, Eqn. 5 simplifies to

$$\overline{N_m} = N_a + \frac{\sigma_\phi^2}{4}, \tag{6}$$

implying a time-averaged null in the absence of an astronomical null leakage signal of one quarter of the phase variance. Indeed, in the special case of phase-only errors being considered at the moment, the astrophysical null is actually given by the signal minimum, which occurs at $\phi = 0$. (However, for completeness, we note that if noisy dark or background terms are subtracted, the astrophysical null is no longer exactly at the minimum (Hanot et al. 2011; Mennesson et al. 2011a, 2011b, 2013, 2016)).

We now separate the phase errors into a long-term mean-phase offset from null, $\bar{\phi}$, and the phase fluctuations, $\Delta\phi$, about that mean, via $\phi = \bar{\phi} + \Delta\phi$. From Eqn. 2, the measured null at any time is then

$$N_m = N_a + \frac{(\bar{\phi}+\Delta\phi)^2}{4}, \tag{7}$$

leading to an average null of

$$\overline{N_m} = N_a + \frac{\bar{\phi}^2}{4} + \frac{\sigma_\phi^2}{4}. \tag{8}$$

The astrophysical null depth is therefore not actually given by the average measured null. Indeed, solving for the astrophysical null depth in Eqn. 7 gives

$$N_a = \overline{N_m} - \frac{\sigma_\phi^2}{4} - \frac{\bar{\phi}^2}{4}, \tag{9}$$

which contains two correction terms to the measured mean null depth, the first being one quarter of the phase variance, and the second being one quarter of the square of the mean phase.

As the phase variance is not directly measured, it is better to cast this equation in terms of the directly measurable null-depth variance. The measured null depth variance, $\sigma_N^2 = \overline{(N_m - \overline{N_m})^2}$, can be found by inserting Eqns. 7 and 8 into this definition, and using the fact that $\overline{(\Delta\phi)^4} = 3\sigma_\phi^4$ for Gaussian fluctuations, yielding

$$\sigma_N^2 = \frac{\sigma_\phi^4 + 2\sigma_\phi^2 \bar{\phi}^2}{8}. \tag{10}$$

This equation gives the variance of the null depth fluctuations as a function of the rms phase fluctuation level and the mean phase offset (for small phase errors). Eqn. 10 is actually a quadratic equation in the phase variance, $\sigma_\phi^2$, with the solution

$$\sigma_\phi^2 = -\bar{\phi}^2 \pm \sqrt{\bar{\phi}^4 + 8\sigma_N^2}. \tag{11}$$

Inserting this result into Eqn. 9 then gives

$$N_a = \overline{N_m} \mp \frac{\sqrt{\bar{\phi}^4 + 8\sigma_N^2}}{4}. \tag{12}$$

Because null-depth fluctuations are positive-definite, the correct sign is negative, finally giving

$$N_a = \overline{N_m} - \frac{\sqrt{\bar{\phi}^4 + 8\sigma_N^2}}{4}, \tag{13a}$$

i.e., in the case of small phase fluctuations about a mean phase offset, the astrophysical null is actually given by the difference between the measured average null depth and a correction term that depends on both the measured null-depth variance, $\sigma_N^2$, and the mean phase offset. For $\bar{\phi} = 0$ Eqn. 12 simplifies to

$$N_a = \overline{N_m} - \frac{\sigma_N}{\sqrt{2}}, \tag{13b}$$

in which both terms on the right hand side can be determined directly from the null-depth data stream. Note that the presence of $\sigma_N$ in Eqns. 13a and b means that characterization of the null-depth fluctuations is *required* in order to determine the astrophysical null depth in the presence of phase (or more generally, other) fluctuations. While both the average null depth and its variance can be determined from measured null-depth data, obtaining the mean phase needed in the more general Eqn. 13a is not as obvious, because to first order the mean phase term simply adds to the measured astrophysical null depth (e.g., Eqn. 8 for $\sigma_\phi = 0$). It is therefore only in the special case of a zero-mean phase that Eqn. 13b gives $N_a$ directly. In the more typical real-world case of a non zero-mean phase (i.e., with an average phase offset due to a phase setpoint error), one could experimentally reduce $\bar{\phi}$ (perhaps using a second-wavelength fringe tracker), but a residual, albeit reduced, $\bar{\phi}$ would inevitably remain. Even worse, $\bar{\phi}$ would not be expected to remain identical from star to star, and so removing the mean phase contribution by using a calibrator star is not expected to lead to great success.

However, further statistical analysis of the null depth fluctuations can provide the mean phase, because the character of the null depth fluctuations is actually a function of the mean phase. It is easy to see why, by considering illustrative setpoints along a sinusoidal fringe: at null, the null depth changes by exactly the same (positive) amount for phase fluctuation of a given magnitude of either sign, while, at quadrature, the null depth changes in opposite directions for phase fluctuations of opposite sign. For a general phase offset and small phase errors of opposite sign, the null depth changes by different amounts, and generally in opposite directions. It was in fact to take advantage of *all* of the statistical information present within a measured null depth sequence that the nulling self-calibration (NSC) technique was originally conceived (Hanot et al. 2011; Mennesson et al. 2011a, 2011b). In brief, NSC calculates the probability density function (PDF) of the fluctuating null-depth data stream, and fits it with model PDFs that are functions of the relevant error variables, including all of the contributing phase and intensity errors. The shape of the PDF has been explored in detail (Hanot et al. 2010, 2011; Mennesson et al. 2011a, 2011b, 2013, 2016; Serabyn 2020), and so is not elaborated here, but the essential point is that the NSC algorithm can be used to fit all of the instrumental error parameters that contribute to the observed null-depth PDF, thereby allowing for a high-accuracy solution for the astrophysical null depth even in the presence of instrumental fluctuations and mean phase offsets. Indeed, as briefly illustrated at the start of this paragraph, to extract an unknown mean phase, null depth fluctuations are actually necessary, because it is the phase-offset-dependent character of the null-depth fluctuations that reveals the underlying phase offset (Hanot et al. 2010; Hanot et al. 2011; Mennesson et al. 2011a; Mennesson et al. 2016; Serabyn 2020).

As a result, the NSC algorithm can greatly relax an interferometer's phase stabilization requirement, which is nominally given by Eqn. 3 as $\frac{\lambda\sqrt{N}}{\pi}$, or $\sim \lambda/300$ for a $10^{-4}$ null. Indeed, with the NSC algorithm, in practice the main requirement was merely found to be the need to avoid fringe hops, so that in rough terms, staying within $\sim \lambda/4$ of the fringe minimum suffices. This corresponds to relaxing the phasing requirement by roughly two orders of magnitude, while still allowing astrophysical detections well below the average instrumental null level set by the fluctuations. This has been borne out in practice, where the NSC algorithm was found to relax the fringe stabilization requirement by factors as large as 30 to 100 (Mennesson et al. 2011a, Kühn et al. 2015). The NSC algorithm is therefore critical in extending deep nulling observations to wavelengths as short as the NIR.

**2.4. The PFN's reduced error budget.** For deep nulling, it is critical to minimize all of the contributing error terms. For small errors, in a single detector readout, the null-depth leakage at the output of a single-baseline nuller is given by a sum of variance terms (Serabyn 2000; Serabyn et al. 2012), i.e.,

$$N = \frac{1}{4}\left(\overline{\emptyset_g^2} + \overline{\emptyset_a^2} + \overline{\emptyset_t^2} + \overline{\emptyset_d^2} + \overline{\emptyset_p^2} + \overline{\alpha^2} + \overline{\delta^2}\right), \tag{14a}$$

were the first term is the leakage due to the finite size of the star (i.e., the geometric phase variance across the stellar disk), the second is the spatial phase variance due to wavefront errors across the beam aperture, the third is the temporal variance of the piston phase between subapertures within an integration time, the fourth is the dispersive phase variance across the passband, the fifth is the differential retardance between the two polarization states, the sixth is the relative polarization rotation between the two sub-aperture beams, and the last term is due to the amplitude mismatch between beams. (Note that each term in this equation averages over a different parameter). In the PFN, several of these terms are rendered negligible by design: the short PFN baseline guarantees that the first term is negligible for all but the closest main sequence stars (but this term is the stellar diameter signal for large enough stars), the PFN's nearly common-mode optical system limits the polarization rotation and retardance errors, and the use of a SM fiber at the PFN's output eliminates higher-order wavefront errors (beyond tip-tilt) across the subapertures. (Detector noise is absent from Eqn. 14a because for the PFN it is negligible compared to photon noise on bright stars.) Of the remaining terms, a SM fiber does transmit piston phase and amplitude mismatch errors (the latter of which can be reduced by flux balancing), as well as dispersive phase errors, thereby reducing the PFN's error budget to

$$N = \frac{1}{4}\left(\overline{\emptyset_t^2} + \overline{\emptyset_d^2} + \overline{\delta^2}\right), \tag{14b}$$

wherein post-ExAO tip-tilt errors have been converted to amplitude errors by the SM fiber (Mennesson et al. 2002; Wallner et al. 2004). The use of a SM fiber combiner after a common-mode optical system thus leaves only relative piston phase, dispersion, and amplitude and pointing errors to be dealt with.

In the PFN, these remaining error terms are reduced as follows (a detailed instrument description is provided in Appendix 2): Cross-aperture piston phase errors are reduced to ~ 100 nm by the ExAO system, after which the phase residuals are estimated with the NSC algorithm. Pointing errors are also first reduced by the ExAO system, after which an H-band pointing camera on the PFN bench removes post-AO drifts. Amplitude mismatch is reduced by means of lateral shear between our dual-subaperture pupil mask, the pupil image, and the single-mode fiber's acceptance cone. Dispersion is minimized by modifying the relative thicknesses of dielectric plates that the two subaperture beams traverse. Finally, the temporal smoothing of the residual post-ExAO phase jitter (the 3rd term in Eqn. 14a) is kept to a minimum with short (5 ms) exposures and rapid calibration, i.e., rapid switching between the four needed observational states - the interferometrically combined signal, both individual beam intensities, and the dark level (Hanot et al. 2011; Kühn et al. 2015; Mennesson et al. 2016). Of course, a SM fiber can only be used effectively if the coupling of the subaperture beams to the fiber is both stable and high. In the case of the PFN, the ExAO system flattens the sub-aperture wavefronts, allowing for good coupling of the fiber mode to subapertures large enough to provide credible NIR SNR, and the pointing to the fiber is stabilized by the combination of the ExAO system and updates from our H-band camera. Fig. 2 provides an example of the

PFN's resultant single-beam fiber-coupling, showing an on-sky coupling stability of ~ 5% rms. In summary, the PFN reaches high null-depth accuracies with a combination of common-mode optics, ExAO fringe stabilization and fiber-coupling stabilization, dielectric dispersion reduction, SM-fiber beam combination, rapid detector readout, rapid calibration, and the NSC data analysis algorithm.

### 3. Some applications of high-accuracy NIR nulling

In this section, we briefly discuss the two applications of nulling interferometry of relevance to this paper: companion detection inside $\lambda/D$, and stellar diameter measurements using short baselines.

**a) Faint companion detection inside $\lambda/D$** One of the main advantages of nulling interferometry is that it can enable observations of companions to bright stars within the stellar PSF core. The flux from a secondary point source of intensity $I_p$ located at an off-axis angle $\theta_p$ that is transmitted by a sinusoidal fringe pattern centered on the primary star is given by

$$I = I_p \sin^2\left(\frac{kb\theta_p \cos(\alpha_p)}{2}\right), \quad (15)$$

where $\alpha_p$ is the angle between the binary's on-sky separation vector and the on-sky baseline direction. This equation does not include the single-aperture response pattern, which attenuates off-axis fringes, but we note here that with a baseline only 2.27 times the sub-aperture diameter, the PFN's response does not extend much beyond a single strong transmission fringe on either side of the null fringe (Mennesson et al. 2008; Kuhn et al. 2014). For arbitrary $\theta_p$, the null-depth rotation curve of Eqn. 15 will produce a series of harmonics of the angle $\alpha_p$, but for small $\theta_p$ the response curve of Eqn. 15 reduces to

$$I = \frac{I_p(kb\theta_p)^2}{4}\cos^2(\alpha_p). \quad (16)$$

Using the appropriate half-angle formula, the nuller's response to point sources located at off-axis separations small compared to the fringe spacing is then seen to appear exclusively at the 2$^{nd}$ harmonic of the baseline rotation rate:

$$I = \frac{I_p(kb\theta_p)^2}{8}\left(1 + \cos(2\alpha_p)\right). \quad (17)$$

The amplitude of the null-depth rotation curve thus depends on the product of instrumental parameters, $(kb)^2$, with a product of source parameters, $I_p\theta_p^2$, the latter implying a degeneracy between the secondary's intensity and angular offset for small angular separations (Kühn et al. 2015). However, if either $I_p$ or $\theta_p$ is known, the other can be determined. Likewise, the companion's azimuth angle (modulo $\pi$) can be determined from the angle of the peak of the null-depth rotation curve. Note that if the companion's azimuth is known, the full rotation curve is not needed, but the minimum is still needed to be able to subtract any stellar leakage contribution. However, a full rotation curve is needed when searching for unknown companions.

**b) Stellar diameter measurements with reduced baseline lengths.** As we show here, high-accuracy nulling allows a reduction in baseline lengths compared to those typically used in standard LBI visibility measurements of stellar diameters. Short baselines imply wide fringes, which in principle make deep stellar rejection possible. For a single baseline nuller, the astrophysical null depth on a uniform stellar disk of diameter θ* is given by (Serabyn 2000)

$$N^* = \left(\frac{\pi b \theta^*}{4\lambda}\right)^2, \tag{18}$$

which can be recast as

$$\theta^* = \frac{4\lambda\sqrt{N^*}}{\pi b}. \tag{19}$$

(For the PFN this becomes $\theta^* = 166\sqrt{N^*}$ mas at $\lambda = 2.15$ μm.) The stellar diameter measurement accuracy, $\delta\theta^*$, is then differentially related to the null depth measurement accuracy, δN, and the baseline accuracy, δb, via

$$\delta\theta^* = \frac{2\lambda}{\pi b}\frac{\delta N}{\sqrt{N^*}} - \frac{4\lambda\sqrt{N^*}}{\pi}\frac{\delta b}{b^2}, \tag{20}$$

Dividing by Eqn. 19 yields

$$\frac{\delta\theta^*}{\theta^*} = \frac{1}{2}\frac{\delta N}{N^*} - \frac{\delta b}{b}, \tag{21}$$

and multiplying by θ* and inserting N* from Eqn. 18 in the first term yields

$$\delta\theta^* = \frac{8\lambda^2}{\theta^*\pi^2}\frac{\delta N}{b^2} - \theta^*\frac{\delta b}{b} \tag{22}$$

In a cross-aperture nuller, the baseline is well determined, implying that the first term generally dominates in the last three equations. In that case, for a given star (fixed θ*) and a fixed wavelength, the surviving first term in Eqn. 22 implies that the attainable stellar-diameter measurement accuracy is proportional to $\frac{\delta N}{b^2}$, ie., for a given stellar-diameter measurement accuracy, the baseline length needed decreases as the square root of the available null-depth measurement accuracy. This baseline reduction factor can be very significant, with, e.g., ~ $10^{-4}$ null depth accuracies being compatible with baselines ten times shorter than the baselines needed for the same (single-measurement) stellar-diameter accuracies for the more typical LBI null-depth (actually, visibility) accuracies of ~ $10^{-2}$. As already shown with earlier PFN measurements (Mennesson et al. 2011b), this order of magnitude decrease in required baseline lengths enables stellar diameter measurements using large single-aperture telescopes instead of separated-aperture LBI. Indeed, the PFN baseline is almost a factor of two smaller than the original interferometric baseline of Michelson and Pease (1921), and the wavelength is longer, but the PFN can measure stellar diameters an order of magnitude smaller, due to its use of nulling. This is not to say that nulling outperforms long-baseline visibility measurements in terms of ultimate accuracy, especially as LBI can make use of different baseline lengths. Rather, it says that baseline length can be traded judiciously for higher measurement accuracies in some cases. On the other hand, the larger (30 m class) planned single-aperture telescopes will be able to benefit from both longer baselines and higher null depth accuracies. Indeed, while nulling on a 3.4 m baseline has thus far allowed diameter measurements of giant stars, on 30 m class telescopes, NIR nulling will allow (and indeed, not be able to avoid) the measurement of main sequence stellar diameters.

## 4. Demonstration of companion detection by means of baseline rotation on η Peg

The PFN was used to observe the spectroscopic binary η Peg on the night of UT 2015 June 26, over a zenith angle range of 3.5° to 21.5° degrees (i.e., < 1.075 air masses). During the observations, the seeing ranged between 1.5 and 2.3 arcsec. On that date, the η Peg binary had a separation of 30 mas and a position angle of 170° (6th Catalog of visual binary stars; Hartkopf, Mason & Rafferty 2008). No prior measurement of the secondary to primary $K_s$-band flux ratio was found in the literature, but extrapolating from shorter wavelength observations (Hummel et al. 1998), a flux ratio of ~ $10^{-2}$ is expected. The exact ratio depends on the spectral type of the secondary, which shows some variation in the literature. This particular binary was selected because the parameters given above allow simultaneous tests of both the PFN's small-angle and high-contrast capabilities, as η Peg's separation was only about a third of the FWHM of the full telescope's diffraction-limited PSF core ($\approx$ 87 mas), while over a full baseline rotation, the null depths should range from ~ $10^{-2}$ with the PFN baseline perpendicular to the projected binary separation vector on the sky, to 3.7 x $10^{-4}$ with the PFN baseline parallel to the binary separation, the latter leakage being set by primary star's diameter of 3.23 ± 0.07 mas (Nordgren et al. 2001). (Note that a $10^{-4}$ null corresponds to a 10 mag flux reduction.) Observing η Peg over a full 180° of baseline rotation should therefore show a flux drop of more than an order of magnitude as the null fringe is rotated to extinguish both stars.

On the night of the observations, we carried out one to three nulling sequences at each of seven PFN baseline orientations spaced uniformly every 30° across a full 180° end-over-end rotation of the baseline. The angular range was centered on a baseline orientation perpendicular to the known binary separation vector, and so includes measurements at or very near both the maximum and minimum of the expected null-depth rotation curve. Each of the fourteen measured nulling sequences was calibrated from its four-state chopper-wheel data, and null depths and error bars were extracted for each data sequence using the NSC algorithm. At a given baseline orientation, repeatability of the extracted null depth between different scans was typically better than 5 x $10^{-4}$, with only one of the 14 scans showing a much larger deviation, likely attributable to a fringe hop, leading to rejection of that scan. The remaining 13 extracted null depths were then averaged for each baseline position angle to produce average measured null depths at each of the seven observed baseline position angles. Finally, the instrumental null depth floor was subtracted from these values to give the final calibrated astrophysical null depths for η Peg. The instrumental null depth floor was determined from an average of all of our nulling observations covering the same zenith angle range (0 to 21.5 degrees) for the eight stars in a small hot dust survey that we have carried out. These 34 null depths, corrected for stellar diameter leakage, yielded an average null depth of 6.3 ± 0.7 x $10^{-4}$, close to the PFN's laboratory azimuth-averaged $K_s$-band null depth limit of ~ 5 x $10^{-4}$, suggesting an absence of detectable null leakage above the stellar photosphere for these stars, on average, of greater than ~ 1 - 2 x $10^{-4}$. (Note that residual instrumental dispersion at the telescope may also contribute to this slight residue.) The final calibrated nulls making up η Peg's null-depth rotation curve are plotted in Fig. 3, as a function of the baseline position angle on the sky. The calibrated null depths plotted in Fig. 3 range over slightly more than an order of magnitude, from a peak of 5.0 x $10^{-3}$ to a minimum of ~ 3.7 x $10^{-4}$, and follow a generally sinusoidal shape, with a minimum slightly above zero. The rms null-depth error bars for each nulling sequence were also obtained from the NSC algorithm, combined in quadrature for each baseline angle, and then combined in quadrature with the instrumental null-floor rms contribution of 0.7 x $10^{-4}$, giving the error bars shown in Fig. 3.

The observed near-sinusoidal shape of $\eta$ Peg's null-depth rotation curve is consistent with a companion well inside the first constructive fringe maximum, as discussed in section 3 (Eqn. 17). With a known binary separation vector, model null-depth rotation curves can be fitted to $\eta$ Peg's observed null depths with only one free parameter - the flux ratio. The resultant best-fit model to $\eta$ Peg's null-depth rotation curve is shown as the solid curve in Fig 3. The data follow the best-fit model very well, with most deviations from the curve being within the 1-sigma error bars (the reduced $\chi^2$ is 0.49, suggesting that the error bars may be slightly overestimated). The best-fit curve, together with reduced $\chi^2$ vs. flux ratio estimates, yield a secondary to primary flux ratio of 1.08 ± 0.06 x $10^{-2}$, which corresponds to a $K_s$ magnitude difference of 4.92 ± 0.06 mag. For a primary of spectral type G2II-III with Kmag = 0.9 at $\eta$ Peg's distance, this is consistent with a secondary of type A5V, in agreement with the spectral type derived by Hummel et al. (1998).

However, as Fig. 3 also shows, neither the data nor the best fit curve bottoms out at zero. The lowest pair of data points (i.e., the first and the last) are both for baseline orientations parallel to the binary separation vector, in which case the null fringe extinguishes both stars. Both of these data points lie significantly above zero, consistent with a residual leakage contribution from the primary star's diameter. In fitting the primary-to-secondary flux ratio, a constant null depth offset of 3.7 x $10^{-4}$, corresponding to the leakage due to the measured stellar diameter of 3.23 mas (Nordgren et al. 2001), was therefore included in the model curves discussed in the last paragraph. However, as the diameter of the primary's stellar disk is independent of the binary flux ratio, a disk size determination can be made using only the two lowest data points, i.e., those at the baseline angles where both stars are nulled. Averaging these two points yields an average leakage due to the primary's stellar disk of 4.8 ± 1.6 x $10^{-4}$, which translates to a stellar diameter of $3.66^{+0.56}_{-0.68}$ mas. This diameter, measured using a single spatial frequency, is consistent with the LBI measurement of 3.23 ± 0.07 mas (Nordgren et al. 2001), but has larger error bars, likely because the spectrally-dispersed, multi-baseline LBI observations made use of 96 spatial frequencies.

## 5. Summary

The $\eta$ Peg nulling data serve both to demonstrate and delimit the PFN's capabilities, including its ability to freely carry out observations at any baseline orientation, its ability to distinguish companion leakage from diameter leakage in the course of a baseline rotation, its dynamic range and high contrast detection capabilities (several x $10^{-4}$), and its small angle capability (~ 30 mas, or roughly 1/3 $\lambda$/D). Finally, note that as 30 mas is 0.23($\lambda$/b), or 0.93 of the PFN's official IWA (Section 2.1), $\eta$ Peg B is just inside the null fringe's half power point, leaving some leeway for companion detection to yet smaller angles.

Even with its short interferometric baseline, the PFN has thus proven to be a valuable demonstration platform for novel nulling approaches, including here finally a demonstration of the original rotating-nuller concept of Bracewell (1978), novel beam-combiner, data-calibration and data-reduction techniques, such as, e.g., fiber nulling (Wallner et al. 2004; Haguenauer & Serabyn 2006) and the NSC algorithm (Hanot et al. 2011; Mennesson et al. 2011a), and also deep nulling at NIR wavelengths. Perhaps the most important thing to emerge from the PFN project is the realization that stringent hardware-based interferometric phase stability can be traded for a much easier software-based post-acquisition statistical fluctuation analysis. As a result, the NSC technique is now also employed at both the LBTI (Defrere et al. 2016) and the Anglo-Australian Telescope (Lagadec et al. 2018). Eventually, NSC should prove even more useful in the case of space-based MIR nulling-interferometry observations of exoplanets, where the background is low enough that phase-related fluctuations should again dominate.

In general terms, because the astrophysical null is given by the signal minimum in the simple phase-only case (Eqn. 2), the NSC algorithm can be seen as the interferometric "dark fringe" analog of coronagraphic dark-speckle companion-detection techniques that operate at intensity minima within coronagraphic dark holes (Labeyrie 1995). Indeed, the coupling of SM fibers to specific locations within dark holes serves to complete the analogy (Mawet et al. 2017). The two techniques differ mainly in the number of sub-apertures contributing to the dark field, ranging from the number of deformable-mirror elements in the coronagraphic case to a pair of much larger apertures in the single-baseline nulling interferometer case. The dark-fringe and dark-speckle techniques can therefore be viewed as opposite extremes within a single family of measurement techniques.

## 6. Prospects

What further advances are possible? First, we note that the PFN sensitivity was far from optimized, as a rather aged detector, with a read noise of ~ 40 e- rms and relatively slow readouts (5 ms), was employed. Both high noise and temporal smoothing therefore impacted performance. Much better detectors exist (e.g., Atkinson et al. 2018), including those enabling GRAVITY's recent performance (Lacour et al. 2019), that should allow nulling faster than the atmospheric fluctuation timescale. Second, residual dispersion from the AO wavefront sensor dichroic and the atmosphere at high airmasses remained limitations. Fortunately, the effects of dispersion can be mitigated with spectrally-dispersed nulling, in which a number of finer spectral channels are nulled individually. With NSC, the different spectral channels need not be nulled simultaneously, as the different mean phase offsets likely to be present in different channels as a result of dispersion can all be removed by NSC after data acquisition, allowing one to extract the astrophysical null as a function of wavelength. Dispersed nulling together with NSC can therefore relax both the phasing and dispersion requirements. Third, in the laboratory, fiber nulling at $K_s$-band has thus far shown a limit of ~ $5 \times 10^{-4}$. This may be fiber related, perhaps resulting from the bright off-center fringes incident on the fiber cladding in the Fizeau configuration. However, surpassing that null-depth level is possible, as the same Fizeau fiber-nuller configuration has already shown H-band laboratory null depths < $10^{-4}$ (Martin et al. 2006) and HeNe null depths close to $10^{-6}$ (Haguenauer & Serabyn 2006). Moreover, if the limitation proves related to the bright off-axis fringes on the fiber tip, one could instead use a coaxial Michelson combiner in which the bright output is well separated from the nulling output. Indeed, a grating beamcombiner (Martin et al. 2017) can convert input beams arriving at focus in the Fizeau configuration into a single coaxial beam prior to fiber injection, a technique that has already achieved laboratory K-band null depths of $4 \times 10^{-5}$. This approach also has the unique capability of producing achromatic fringes on the sky. With read noise, the readout rate and dispersion all reduced significantly, and the current laboratory NIR null-depth limit overcome, on-sky null depths could be expected to improve by an order of magnitude.

Improved nulling performance can be put to good use on the larger telescopes of the future, where both longer baselines and larger numbers of baselines are possible. This can be taken advantage of in a variety of ways, including simultaneously acquiring single-baseline nulling data over a range of baseline lengths and/or orientations, the latter potentially eliminating the need for physical baseline rotation. Larger numbers of baselines can also be used to provide deeper rejection, using either "higher-order" nulling configurations that combine more than two apertures into a single common null (Angel & Woolf 1997; Mennesson & Mariotti 1997; Velusamy, Beichman & Shao 1999; Karlsson et al. 2004; Serabyn 2004), or phase chopping between different nulling baselines (Beichman, Woolf & Lindensmith 1999; Absil, Karlsson & Kaltenegger 2003; Mennesson, Leger & Ollivier 2005). Indeed, because longer nulling baselines provide more modest stellar rejection (Eqn. 18), more complex NIR nulling configurations may be necessary on very large telescopes. In space, additional beams imply additional telescopes, but combining multiple subapertures originating within a single large ground-based telescope pupil requires only a more complex beam combiner, making the use of multiple nulling baselines more straightforward in the ground-based case. Small subapertures would maintain a large field of view, while larger subapertures could aim at ultimate sensitivity and longer wavelengths, a bifurcation perhaps naturally suited to the pupils of the Thirty Meter Telescope (TMT) and Extremely Large Telescope (ELT) on the one hand, and the Giant Magellan Telescope (GMT) on the other. Ultimately, an entire telescope pupil could be paved with subapertures, as in some versions of aperture masking interferometry (e.g., Lagadec et al. 2018) and the general coronagraphic concepts described by Guyon et al. (2006) and Serabyn (2009). Using such sub-aperturing, one could potentially begin very small-angle high-contrast surveys by nulling between large subapertures to maximize the signal-to-noise ratio and to be sensitive to both symmetric and asymmetric off-axis emission, and then further characterizing any asymmetric emission with closure-phase aperture masking. However, nulling would remain essential to seeing the overall background disk emission and in distinguishing between point sources and more extended disk hot spots.

Finally, we note that implementing a cross-aperture nuller on a large telescope in the future may not require an additional instrument, if the telescope is already equiped with a high-contrast coronagraphic bench. Coronagraphs usually include at least one focal plane and an associated downstream pupil (Lyot) plane wherein optical masks can be inserted, so a nuller could potentially be implemented simply by inserting appropriate "nulling masks" into these planes. For example, a grating nuller (Martin et al. 2017) would require only a grating in the coronagraphic focal plane and a central subaperture mask (~ 1/3 the pupil image size) in the Lyot plane. Fiber ports also already exist behind some high-contrast coronagraphic benches (Jovanovic et al. 2018), but nulling would require that fiber to be single mode. With a nuller implemented simply by adding additional masks to an existing coronagraph, nulling interferometers may thus evolve to simply become additional small-angle modes of high-contrast coronagraphs.

What observations can cross-aperture nulling on large telescopes enable? NIR nullers on large telescopes can provide unique access to high-contrast observations at angles well inside the coronagraphic regime. For example, one could search for brown dwarfs, including the brightest of the "long-term-trend" RV candidates (Crepp et al. 2016; Ryu et al. 2016), up to an order of magnitude closer to stars than coronagraphs can typically reach (in to ~ 0.2 λ/D vs. 1-3 λ/D). Nullers could also probe the mysterious inner dust regions seen around nearby A stars such as Vega (Absil et al. 2013). Indeed, for Vega, maximal dust temperatures and existing PFN observations together constrain any hot dust to lie between 0.1 – 0.2 AU, or 13 – 26 mas (Mennesson et al. 2011b), which a NIR nulling baseline only a factor of two longer than the PFN should be able to reach. Such baseline lengths are already feasible at existing 8 – 10 m telescopes or the LBTI. Moreover, cross-aperture nullers on 30 - 40 m telescopes, with H- and K-band nulling inner working angles on the order of 2 – 4 mas, can potentially provide direct imaging and spectroscopic access to the innermost Hot Jupiters, at radial offsets of a few mas and contrasts of ~ $10^{-4}$ to $10^{-5}$. Thus, with Hot Jupiters, inner exozodiacal dust regions, young protoplanets embedded in disks, and massive long-term-trend RV candidates reachable, NIR nullers on large telescopes should be able to access a significant set of interesting source types, with the accessible region reaching inward to a few mas with the next generation of large telescopes.

**Appendix 1. Why null?**

To answer this question, we consider the interferometric noise versus fringe phase. For a nuller, the intensity, I, obtained upon interfering two monochromatic beams of intensities $I_1$ and $I_2$ is given by

$$I = \tfrac{1}{2}\left(I_1 + I_2 - 2\sqrt{I_1 I_2}\cos(\phi)\right), \tag{A1}$$

where $\phi$ is the phase difference between the two beams relative to the null phase of $\pi$ radians. We now consider the case of small phase and intensity fluctuations about various setpoints, with a given setpoint being defined by the mean phase $\bar{\phi}$ and by equal beam intensities, i.e., $I_2 = I_1$. Allowing for fluctuations, at any time we have as before that $\phi = \bar{\phi} + \Delta\phi$, and now also that $I_2 = I_1 + \Delta I$. We then have

$$I = I_1 + \tfrac{\Delta I}{2} - I_1\sqrt{1 + \tfrac{\Delta I}{I_1}}\cos(\bar{\phi} + \Delta\phi). \tag{A2}$$

Expanding the square root for small ΔI, we find

$$I = I_1 + \tfrac{\Delta I}{2} - I_1\left(1 + \tfrac{\Delta I}{2I_1} - \tfrac{1}{8}\left(\tfrac{\Delta I}{I_1}\right)^2\right)\left(\cos(\bar{\phi})\cos(\Delta\phi) - \sin(\bar{\phi})\sin(\Delta\phi)\right). \tag{A3}$$

Now expanding the small sinusoidal terms, we have

$$I = I_1 + \tfrac{\Delta I}{2} - I_1\left(1 + \tfrac{\Delta I}{2I_1} - \tfrac{1}{8}\left(\tfrac{\Delta I}{I_1}\right)^2\right)\left(\cos(\bar{\phi}) - (\Delta\phi)\sin(\bar{\phi}) - \tfrac{(\Delta\phi)^2}{2}\cos(\bar{\phi})\right). \tag{A4}$$

Multiplying out the two terms in parentheses, keeping error terms only up to 2$^{nd}$ order, and rearranging, we get

$$I = I_1\bigl(1 - \cos(\bar{\phi})\bigr) + \tfrac{(1-\cos(\bar{\phi}))}{2}\Delta I + I_1\sin(\bar{\phi})(\Delta\phi) + \tfrac{I_1\cos(\bar{\phi})}{2}(\Delta\phi)^2 + \tfrac{\cos(\bar{\phi})}{8I_1}(\Delta I)^2 + \tfrac{\sin(\bar{\phi})}{2}(\Delta\phi)(\Delta I). \tag{A5}$$

At any operating point at $\bar{\phi}$ and $I_2 = I_1$, this simplifies to

$$I_{op} = I_1\bigl(1 - \cos(\bar{\phi})\bigr), \tag{A6}$$

and the intensity fluctuations about that operating point are then given by

$$I - I_{op} = \frac{(1-\cos(\bar{\phi}))}{2}\Delta I + I_1 \sin(\bar{\phi})(\Delta\phi) + \frac{I_1 \cos(\bar{\phi})}{2}(\Delta\phi)^2 + \frac{\cos(\bar{\phi})}{8 I_1}(\Delta I)^2 + \frac{\sin(\bar{\phi})}{2}(\Delta\phi)(\Delta I). \quad (A7)$$

The first two of these terms are linear in the phase and intensity errors, while the last three are quadratic.

Three locations of the operating point are illustrative, i.e., the fringe maximum (or peak) at $\bar{\phi} = \pi$ (where the intensity is $I_p = 2 I_1$), the half-power (or quadrature) point at $\bar{\phi} = \frac{\pi}{2}$, and the minimum (or null) at $\bar{\phi} = 0$. In these cases, we have

Peak: $$I - I_p = \Delta I - I_p\left(\left(\frac{\Delta\phi}{2}\right)^2 + \left(\frac{\Delta I}{2 I_p}\right)^2\right) \quad (A8a)$$

Quadrature: $$I - I_q = \frac{1}{2}\left(\Delta I + I_p(\Delta\phi) + (\Delta I)(\Delta\phi)\right) \quad (A8b)$$

Null: $$I - I_n = I_p\left(\left(\frac{\Delta\phi}{2}\right)^2 + \left(\frac{\Delta I}{2 I_p}\right)^2\right) \quad (A8c)$$

In these equations, error terms linear in phase and/or amplitude are present at both peak and quadrature, but all linear error terms vanish at null, leaving only smaller quadratic error terms. Operating at the null phase therefore yields the minimum noise, and the inclusion of data from any other fringe phases (which a standard full-fringe visibility measurement would necessarily entail) would yield higher noise. The calibration procedure must also avoid higher-noise fringe phases, and so the PFN calibrates by means of individual single-beam intensities, which are not affected by interferometric noise. Due to the reduced noise at null, nulling interferometry thus has the potential to reach deeper contrasts than any visibility-based interferometry technique that makes use of multiple fringe-phase data. However, this conclusion does not apply in the background-dominated case (i.e., the thermal infrared), in which all fringe phases are equally beset by the same high background noise level. Therefore, in the ground-based case, nulling interferometry has the potential to outperform standard interferometry mainly at the shorter wavelengths (i.e., in the NIR and potentially the visible), as long as one is in the source-photon-dominated noise regime.

**Appendix 2. The final architecture of the Palomar Fiber Nuller**

**Configuration overview.** The PFN's optical system has been described previously (Mennesson et al. 2006; Serabyn and Mennesson 2006; Martin et al. 2008; Mennesson et al. 2010; Serabyn et al. 2010; Kühn et al. 2014), but several upgrades have taken place over the years, and so a brief overview of the final system configuration is provided here, and in Fig. 4. The PFN combines the $K_s$-band light from a pair of symmetrically-located pupil subapertures in a Fizeau configuration, with the two subaperture beams interfering in a common focus on the tip of a SM fiber (Wallner et al. 2004; Haguenauer & Serabyn 2006). The PFN's optical system is composed of two subsystems: the nulling optical bench (which includes all of the optical functions up to and including fiber injection), which is mounted at the exit of the Palomar ExAO system in the telescope's Cassegrain cage, and a cryogenic infrared camera located on a movable cart on the dome floor that the nulled light reaches through a 20-m-long infrared single-mode fiber.

**The AO bench.** The telescope beam first traverses the Palomar AO bench (Dekany et al. 2013), where a special dichroic transmits 20% of the visible light and 100% of the J, H and K passbands to the PFN bench, while reflecting 80% of the visible light to the wavefront sensor. Because the tilted AO wavefront sensor dichroic lies in a converging beam, it introduces angularly dependent dispersion, which is compensated by a tilted flat optic located at the entrance to the nulling bench (Kühn et al. 2014).

**Pupil rotation and beam shear.** Once on the PFN bench (Kühn et al. 2014), the beam is collimated by an off-axis paraboloidal mirror, combined with a 1550 nm alignment laser (injected thru a central hole in a flat fold mirror), and sent through a K-mirror that can be used to rotate the telescope pupil about its center. The light then reflects off a pair of mirrors making up a right-angle periscope that can be used to shear the beam in both lateral directions (and to correct residual K-mirror tip-tilt errors). The beam next encounters a dichroic splitter that transmits the J and H band light toward a NIR InGaAs pointing/tracking camera (which operates at H-band, as the J-band light is blocked by a filter), and reflects the $K_s$-band light and the residual visible light onward to the nuller.

**Subaperture and baseline definition.** The reflected beam next encounters a dual-aperture pupil mask located near an image of the primary mirror. A selection of aperture pairs located symmetrically across the telescope pupil image is available on a rotatable wheel (Fig. 4), but we typically used the aperture pair with largest elliptical subapertures that can fit between the outer edges of the primary mirror and the blockage caused by the central secondary-mirror support, for which the baseline length between sub-aperture centers, as imaged onto the primary mirror, is 3.4 m. After selecting an aperture pair, the subaperture mask is fixed for subsequent observations in order to fix the sub-aperture beam footprints on the downstream optics. On the other hand, the sub-aperture baseline vector projected back onto the primary mirror can be rotated about the telescope's optical axis with the K-mirror. The relationship between the K-mirror rotation angle and the on-sky baseline position angle was determined to an accuracy of ±1° using H-band images of known binary stars obtained at several K-mirror rotations.

**Optical path difference.** The two subaperture beams next reflect off of an adjacent pair of flat mirrors (the "split mirror"; Fig. 4) each of which has three-axis piezoelectric actuation that can be used for longitudinal piston offsets between the two beams and individual mirror tip-tilt control. Both mirror pistons are sensed capacitatively, while the tip-tilt angles of both elements are sensed via laser reflections to a pair of quad cells.

**Calibration wheel.** The beams next traverse a dual-slotted chopper wheel (Fig. 4). The chopper wheel's offset elongated apertures allow for rapid switching between the four needed signal states. The final wheel rotation rate was generally ~ 2.5 Hz, and half the rotation period was used to measure the combined beam, while one sixth of the period was used to measure each of the two individual beam intensities and the dark level, respectively. The camera read time was 5 ms, so after blanking off transition regions between states, approximately a dozen useful individual reads were measured in each of the single beam and dark states per wheel rotation, and over three times that many on the nulled state.

**Dispersion corrector and beam compressor.** A visible/IR dichroic next transmits the remaining visible light to a pupil-viewing camera and reflects the $K_s$-band beam pair to a chevron-shaped dielectric element (Fig.

4) that can be rotated about an axis normal to the chevron shape to change the relative dielectric thicknesses seen by the two beams. This dielectric element is made of highly-homogeneous Heraeus Infrasil 301 glass, and it has two roles: its chevron shape allows the two beams to be compressed laterally for more efficient injection (~ 29%) into the SM nulling fiber located further downstream, and it can be rotated about an axis normal to the chevron shape to modify the differential glass thickness seen by the two beams.

**Nuller.** Finally, an off-axis paraboloidal mirror focuses both of the sub-aperture beams onto the face of a SM fiber (a Thorlabs SM2000 fiber with cutoffs of 1.7 and 2.3 microns), where beam combination occurs. With a relative $\pi$-radian phase shift (see nulling paragraph below) between the two arriving beams, the starlight produces an anti-symmetric field distribution on the fiber's entrance plane that cannot propagate in the fiber's single spatial mode (Wallner et al. 2004; Haguenauer and Serabyn 2006). On the other hand, light from off-axis sources that arrives with different relative phases can propagate down the fiber. The resultant angular response pattern on the sky can be seen in Fig. 1 of Mennesson et al. (2010) and Kühn et al. (2014). The PFN's multi-axial beam configuration is more sensitive to pointing errors than coaxial combination (Wallner et al. 2004), with null depths of $10^{-3}$ to $10^{-4}$ corresponding to pointing errors of roughly 1.2 to 0.4 mas, respectively. However, pointing fluctuations (which are in effect phase fluctuations) merely yield additional null depth fluctuations, which are already dealt with by the NSC algorithm. The actual pointing stability needed is therefore defined simply by the need to avoid fringe hops, which means keeping the null-fringe centered on the fiber tip to about a quarter of a fringe, as discussed earlier (Section 2.3).

**The PFN camera.** The 20-m long SM fiber transports any residual non-nulled $K_s$-band light to the infrared detector, which is one of the fringe tracker cameras of the erstwhile Palomar Testbed Interferometer (Colavita et al. 1999). Upon exiting the SM fiber, the light is recollimated, passed through a set of selectable filters and/or attenuators (the latter only for very bright stars), and focused onto the detector array by an external lens. The passband is defined by a cold $K_s$ filter inside the cryogenic dewar, but is impacted by additional spectral rolloffs to long wavelengths (including that of the SM fiber), leaving the band center at ~ 2.15 µm.

**System symmetry and stability.** Because the sub-aperture beams propagate side-by-side through the AO and PFN optical benches, impinging on common optical elements everywhere except for the independent halves of the split mirror, the optical system is essentially common-mode, resulting in a high degree of symmetry between the beam pair, as well as low vibration sensitivity. The PFN bench is also enclosed with covers to reduce air currents.

**Alignment.** Alignment information is obtained with the PFN bench's H-band and visible cameras, both of which can be switched between observing pointing and beam shear. The H-band camera can monitor either the pointing of H-band stellar images or pupil shear relative to the fiducial alignment-laser spot, while the visible camera can show either pupil shear or stellar images for both subapertures. After each move to a new star and each baseline rotation, the stellar boresight on the PFN's H-band camera must be restored, as the K-mirror's optical axis is not perfectly parallel to the chief ray. To accomplish this, the alignment laser is turned on and its pointing on the H-band camera is restored by tip/tilting the periscope mirrors. With the H-band camera in pupil mode, the beam shear is then restored by translating the periscope mirrors. Once satisfactory, the laser is turned off and the camera is switched back to view the target star, which now appears inside the target alignment box. To remove any slow pointing drifts between the PFN and AO benches as the telescope tracks, slow updates are sent to the AO system based on the H-band stellar positions. Finally, we optimize the coupling of the $K_s$-band beam pair from the star to the SM fiber by means of spiral searches carried out using the tip-tilt capabilities of the two split mirror elements in sequence. The pointing of the split-mirror elements is then maintained by a fast control loop using our pair of internal alignment lasers and quad cells (Fig. 4).

**Nulling.** To null starlight, the electric fields of the two subaperture beams must be brought into anti-phase. To this end, a static achromatic $\pi$ phase shift is introduced between the two beams by the combination of a split-mirror pathlength offset and a chevron differential glass thickness. This approach is completely analogous to that used by the Keck Interferometer Nuller (Mennesson et al. 2003, Koresko et al. 2003), except that the dielectric thickness is modified by rotation rather than translation, as described in Martin et al (2003). (For example, for a bandwidth of 0.4 μm, air and dielectric offsets of 135 μm and 123 μm, respectively, would be needed, with an accuracy of ~ 7 μm.) The achromaticity of the fringes is optimized by symmetrizing the depths of the fringes to either side of the central null fringe (Martin et al. 2003), using long scans carried out by the split mirror piezoelectric piston stages. From our modeling, a null depth of ~ $10^{-4}$ requires the two neighboring fringes to be balanced to about 10% of their average intensity.

**Observing checks and data acquisition.** Finally, we note that the Palomar ExAO system can lock onto a given fringe to ~ 100 nm only under good seeing conditions. As a result, the central fringe must be checked at the start of each observing sequence, i.e., for each baseline orientation. To this end, long fringe scans are carried out using the split mirror's piezoelectric actuation. Once the interferometer is phased properly on the central null, sequences of nulling and calibration data (both individual beam intensities and the dark level) are taken through the spinning chopper wheel. Typically, several one- to three-minute long nulling sequences are recorded, with repeats often taken to guarantee that occasional fringe hops can be identified and discarded. During observations, both the raw null depth measurement sequences and their frequency histograms are plotted in real time on the instrument console to judge data quality.

**Acknowledgements** This work was carried out at the Jet Propulsion Laboratory, California Institute of Tedchnology, under contract with NASA. We gratefully thank the staff of the Palomar Observatory and Frank Loya for their assistance, and the referee for many questions that led to an improved paper.

**Figures**

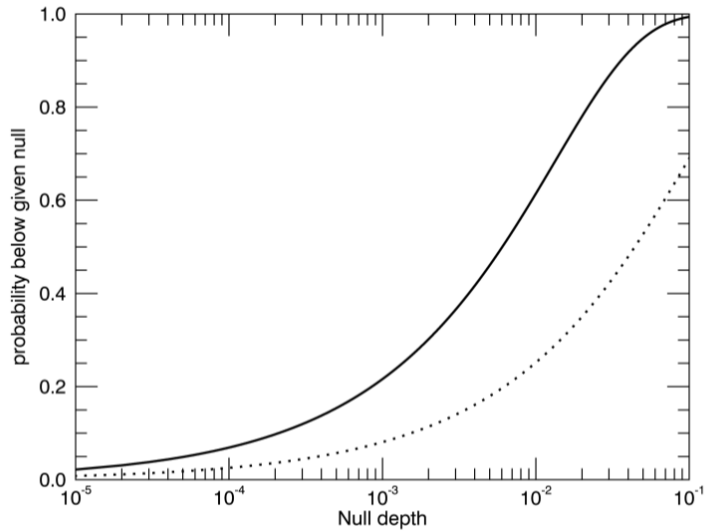

Fig. 1. Time fraction (probability) that instantaneous nulls depths are below a given value. The dotted curve is representative of first generation AO wavefront-error correction levels (Strehl ratio = 0.6), and the solid curve is representative of ExAO wavefront-error correction levels (Strehl ratio = 0.95).

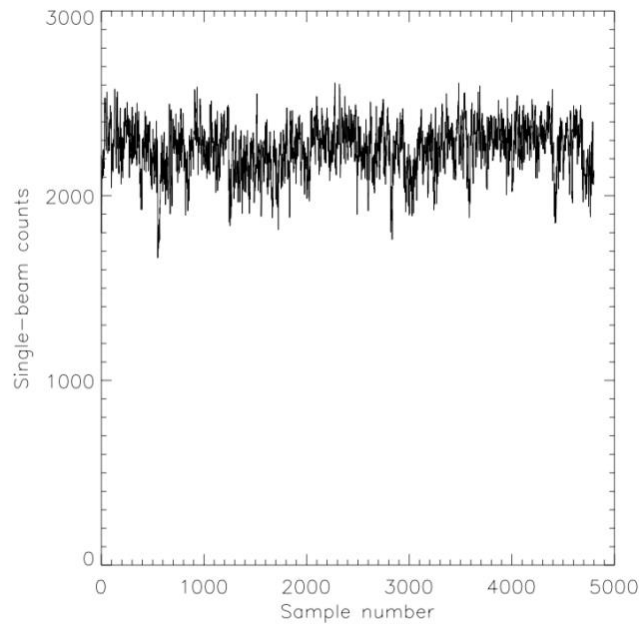

Fig. 2. Single beam signal from zeta Aql (K_mag = 2.9) on 2015 June 28, showing the stability of the stellar signal injected into our SM fiber (under 1.4" seeing conditions). This data was taken during a normal observing sequence, alternating between the two individual beams, the dark and the combined beam, but the other three states have been removed here. Each sample was 5 ms long, implying 25 sec of single beam data. Allowing for gaps due to measurements of the other three states, the entire duration of this data set was roughly 2.5 min. The rms level is 5.2% of the average signal level.

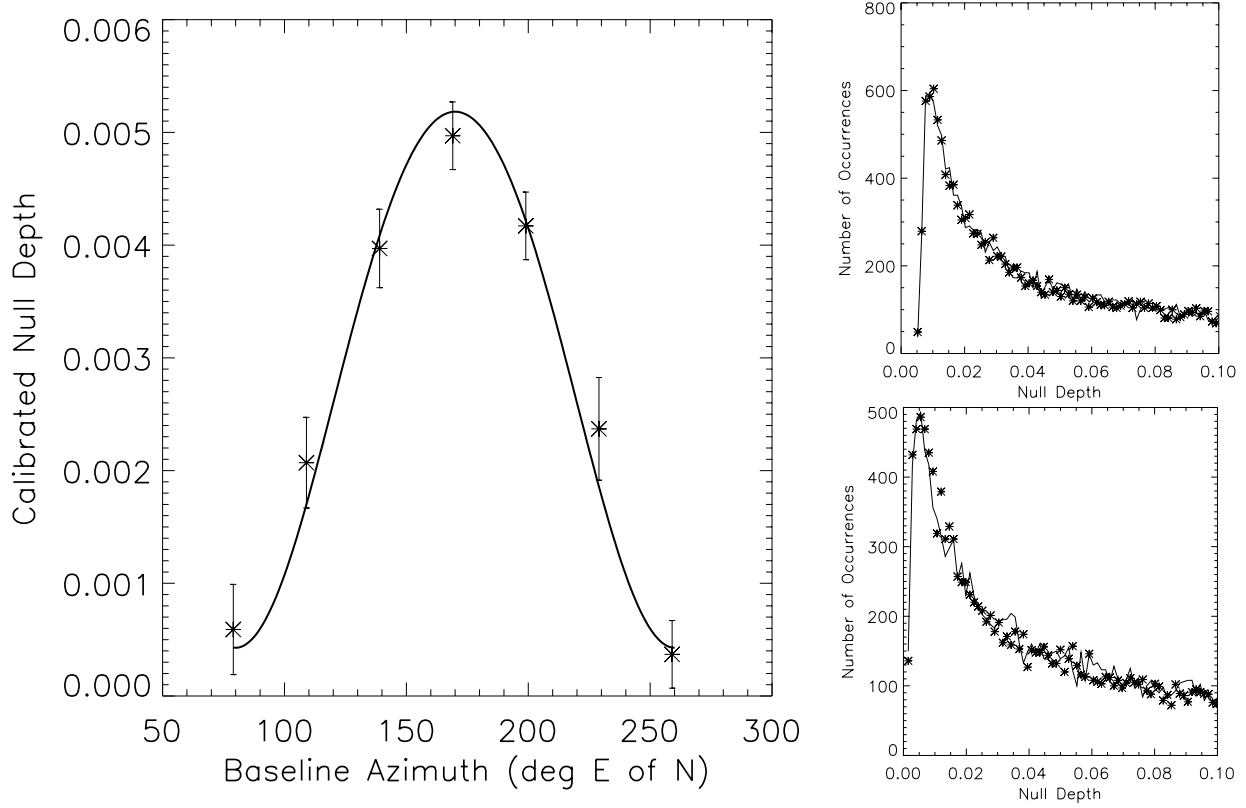

Fig. 3. Left: Calibrated null depths measured on η Peg vs. baseline position angle on the sky over a full 180° of baseline rotation. Right: individual raw (uncalibrated) null-depth histograms corresponding to baseline orientations of, top right: 170° (the observed null-depth maximum), and, bottom right: 260° (the observed null-depth minimum). The shift in peak position reflects the different null depths.

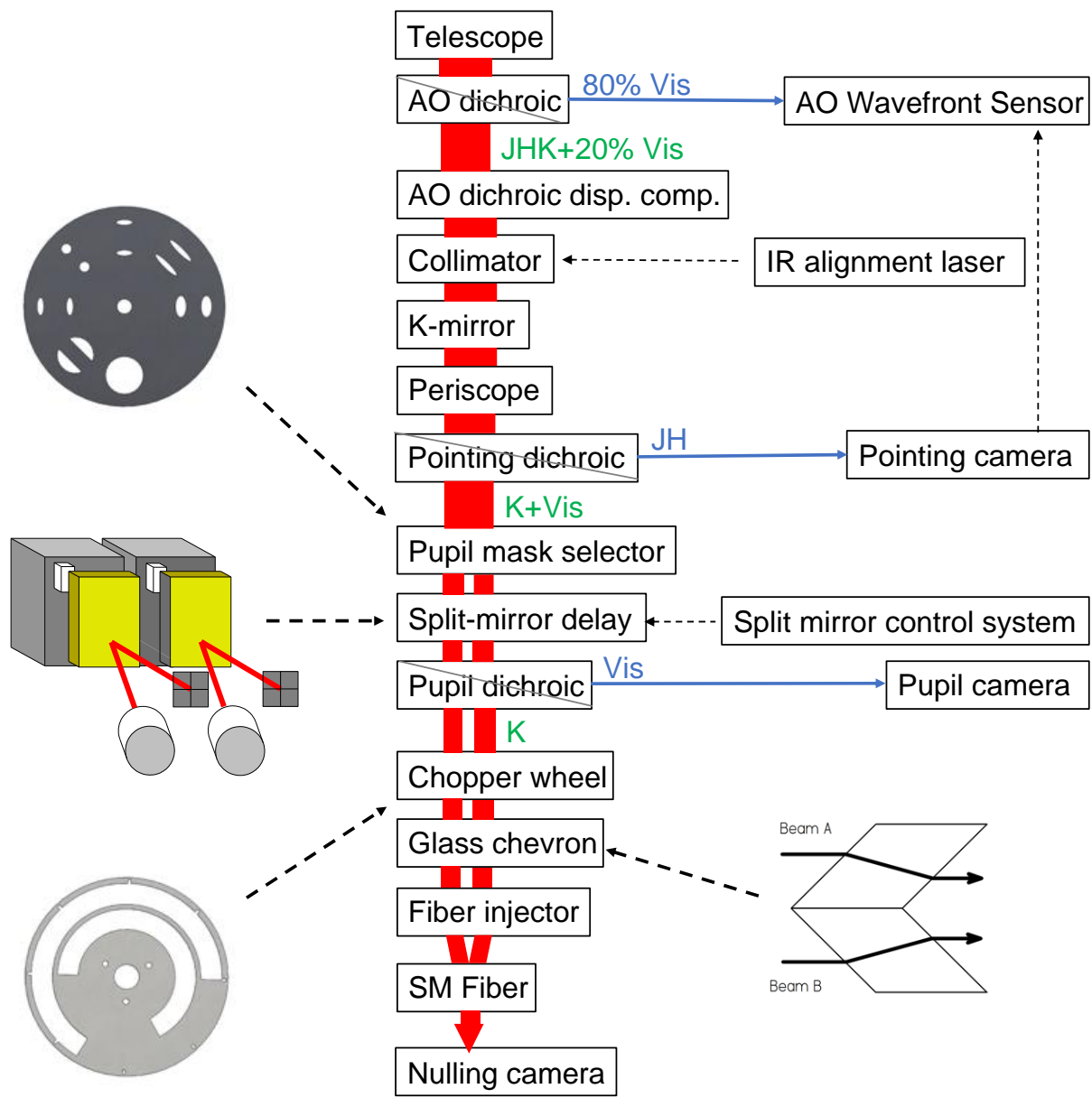

Fig. 4. Block diagram of the Palomar Fiber Nuller's final configuration. The large red arrows trace the flow of the $K_s$-band light, while the thin blue arrows follow the other wavelengths as they split off to other cameras. (The J-band starlight is not used, being filtered out prior to reaching the pointing camera.) Insets along the sides show: Top left: The subaperture mask, with our favored elliptical apertures seen at 3 o'clock. Center-left: The split mirror, with its (laser/quad-cell) tip-tilt sensors and (capacitative) piston sensors shown graphically in front of and behind the (yellow) mirrors, respectively. Bottom left: The final dual-slot/dual-beam chopper wheel. Bottom right: the glass chevron used for dispersion correction and beam compression. Also indicated as light dashed arrows are the split-mirror control system (piston and tip-tilt), and the pointing updates sent from the PFN's pointing camera to the ExAO system.